\documentclass[conference]{IEEEtran}
\IEEEoverridecommandlockouts
\usepackage{enumerate}
\usepackage{cite}
\usepackage[hang,flushmargin,bottom]{footmisc} 

\usepackage{graphicx}
\usepackage{tikz}
\usepackage{pgfplots}
\usepackage{pgfplotstable}
\usetikzlibrary{patterns}
\pgfplotsset{compat=1.16}
\graphicspath{ {./figures/} }
\usepackage{booktabs}
\usepackage{url}
\usepackage{atbegshi,picture}

\usepackage{multirow}
\usepackage{array, makecell}
\usepackage{multicol}
\usepackage{hhline}
\usepackage{colortbl}
\usepackage{caption}
\newcolumntype{R}[1]{>{\raggedleft\arraybackslash}p{#1}}
\newcolumntype{M}[1]{>{\centering\arraybackslash}m{#1}}
\usepackage{algpseudocode}
\usepackage{algorithm}
\usepackage{amsmath}
\usepackage{amsfonts}

\AtBeginShipoutNext{\AtBeginShipoutUpperLeft{%
  \put(\dimexpr\paperwidth-3cm\relax,-1.5cm){\makebox[0pt][r]{Accepted for publication in International Conference on Machine Learning  and Cybernetics (ICMLC) 2023}}%
}}

\begin{document}
\title{A Testbed for Automating and Analysing Mobile Devices and their Applications \thanks{
The Commonwealth of Australia (represented by the Defence Science and Technology Group) supports this research through a Defence Science Partnerships agreement. Lachlan Simpson is supported by a scholarship from the University of Adelaide.}}

\author{\IEEEauthorblockN{Lachlan Simpson\IEEEauthorrefmark{1}, Kyle Millar\IEEEauthorrefmark{2},
Adriel Cheng\IEEEauthorrefmark{1}$^{,}$\IEEEauthorrefmark{2}, Hong Gunn Chew\IEEEauthorrefmark{1}, and
Cheng-Chew Lim\IEEEauthorrefmark{1}}
\IEEEauthorblockA{
\IEEEauthorrefmark{1}School of Electrical and Mechanical Engineering, The University of Adelaide, Australia\\
\IEEEauthorrefmark{2}Information Sciences Division, Defence Science \& Technology Group, Australia \\
Email: \{lachlan.simpson, adriel.cheng, honggunn.chew, cheng.lim\}@adelaide.edu.au\IEEEauthorrefmark{1} \\
Email: \{adriel.cheng, kyle.millar1\}@defence.gov.au\IEEEauthorrefmark{2}
}}
\maketitle

\begin{abstract}
 The need for improved network situational awareness has been highlighted by the growing complexity and severity of cyber-attacks. Mobile phones pose a significant risk to network situational awareness due to their dynamic behaviour and lack of visibility on a network. Machine learning techniques enhance situational awareness by providing administrators insight into the devices and activities which form their network. Developing machine learning techniques for situational awareness requires a testbed to generate and label network traffic. Current testbeds, however, are unable to automate the generation and labelling of realistic network traffic. To address this, we describe a testbed which automates applications on mobile devices to generate and label realistic traffic. From this testbed, two labelled datasets of network traffic have been created. We provide an analysis of the testbed automation reliability and benchmark the datasets for the task of application classification. 
\end{abstract}
\bigskip

\section{Introduction}
The increase in cyber-attacks and other malicious activity within networks has highlighted the need for enhanced network situational awareness. Classifying flows\footnote{We adopt the standard convention of a flow as the sequence of packets with the same 5-tuple (source IP, destination IP, source port, destination port, transport layer protocol)\cite{b29}.} of network traffic aids situational awareness by providing network administrators insight into the devices and traffic on their network. Bring your own device polices and new encryption methods have exacerbated the complexity of traffic classification. Mobile phones are a prominent example of a device that has contributed to the complexity of traffic classification. 

Machine learning (ML) techniques have shown high performance and the ability to handle variants of network traffic\cite{b17,b31,b26,b27}. Machine learning has therefore emerged as the predominant method for mobile phone traffic classification \cite{b8}. To effectively classify the traffic of mobile phones, ML models require extensive network traffic data with reliable ground-truth\cite{b16,b19}. The network traffic of users sourced from operational networks, however, contains sensitive user information and proprietary data. Instead, network traffic testbeds which create and collect network traffic offer a solution. Network traffic testbeds are experimental or automated. 

Experimental testbeds collect data in two ways: (i) over a fixed time interval, one or more applications are manually used to generate network traffic, or (ii) programs are deployed on a mobile device to generate and gather the data whilst maintaining privacy. Manual network traffic generation and deploying programs on mobile devices is infeasible to scale to a large range of applications and devices. Automated testbeds, in contrast implement a software automation framework with user interface (UI) fuzzing. UI fuzzing refers to supplying the mobile application with random inputs to simulate application use. UI fuzzing is limited as it is difficult to determine if randomly interacting with an application produces realistic traffic. 

Whilst experimental and automated testbeds do not violate user privacy, they fail to capture and label background applications. Experimental and automated testbeds fail to capture and label background applications\footnote{Applications that are operational but are not currently being interacted with by the user.}. Testbeds are also currently purpose-built for specific research goals, targeting particular applications and devices. Purpose-built testbeds limit the ability of the testbed to scale to new devices, applications, and different areas of research. The rapid deployment of new applications and devices for consumers necessitates a testbed which can generate realistic data for traffic classification models.

The contribution of this paper is three-fold:

\begin{enumerate}
    \item A novel testbed that provides labelled application traffic of both Android and iOS mobile devices is presented.
    
    \item An automation framework that simulates realistic user behaviour on a mobile phone has been developed. The framework thus improves the realism and repeatability of the testbed over those that utilise UI fuzzing. 

    \item Two datasets \footnote{https://github.com/lachieS/Testbed-Captures.} of 40-hour network captures that were generated on our testbed environment are provided. Analysis of these datasets and their benchmarks for application classification has been carried out. These benchmarks will facilitate future traffic classifier research for network situational awareness.
\end{enumerate}

The remainder of this paper is structured as follows: Section II provides a survey of related work. Section III introduces the testbed. Section IV provides an analysis of the dataset generated by our testbed. Section V discusses the reliability of the automation process. In Section VI we benchmark three ML classifiers on the dataset. Section VII discusses the limitations of the testbed. We conclude in Section VIII with a discussion for future work.

\section{Related Work}
In this section, we survey the current capabilities of existing testbeds and datasets for application classification of mobile devices.
\subsection{Testbeds}
Several testbeds exist for collecting data from mobile devices. The data collected involves battery usage, location information, network protocol performance, and user behaviour patterns \cite{b1}. Current testbeds can be separated into two streams.

\begin{enumerate}
\item \textbf{Real-World Users:} Examples of testbeds of real-world users include, PhoneLab \cite{b2}, NetSense \cite{b3}, and Sensibility \cite{b4}. These testbeds deploy programs only on Android phones used by real-word users to monitor and gather data from sensors and applications. Testbeds of real-world users benefit from collecting realistic data but present several limitations. Capturing specific network traffic relies on the assumption that users are actively performing a certain action. Real-world user testbeds therefore lack data reproducibility. Sensitive user information and experimental time constraints are additional limitations introduced by real-world users.

\item \textbf{Automated:} A crucial benefit of automated testbeds, which real-world testbeds lack, is the ability to repeat controlled experiments. Examples of automated testbeds include Triangle \cite{b5}, MobiBed \cite{b6} and the work presented in \cite{b20,b25}. Triangle focuses on the performance of applications and devices on a 5G network, the work presented in \cite{b20,b25} test IoT vulnerabilities, and MobiBed tests network protocols.
\end{enumerate}

Table \ref{table:testbed summary} summarises the capabilities of current testbeds and where our testbed differs. Our testbed is the first to automate a wide range of realistic user actions on a diverse set of applications and operating systems. Furthermore, our testbed is the first to produce labelled datasets of both iOS and Android mobile device traffic.

\begin{table*}[t]
\centering
\caption{Summary of current testbeds.}
\begin{tabular}[t]{llllll}
\toprule
Testbed & Real-World Users & OS & Automated Applications & Research Area & Available Dataset\\
\midrule
PhoneLab \cite{b2} & Yes & Android & No & Device Usage & No\\
NetSense \cite{b3} & Yes & Android & No & User Behaviour/Social Interactions & No\\
Sensibility \cite{b4} & Yes & Android & No & Mobile Device Sensors & No\\
Triangle \cite{b5} & No & Android/iOS & Yes & Application/Device Performance & No\\
MobiBed \cite{b6} & No & Android & No & Network Protocols & No\\
\cite{b20, b25} & No & Android & No & IoT device vulnerabilities & No\\
\cite{b7} & Yes & Android & No & User Privacy & No\\
Our testbed & No & Android/iOS & Yes & Application/OS Classification and Analysis & Yes\\
\bottomrule
\end{tabular}
\label{table:testbed summary}
\end{table*}

\subsection{Datasets}
We review the features, applications and data generation processes used in datasets for mobile traffic classification research. The common features include time-based flow information such as the time between flows, frequency of packets, and number of bytes per flow\cite{b17,b10,b12,b13,b14}. However, \cite{b9} uses payload signatures to classify applications rather than time-based flow features. The datasets in \cite{b17, b12,b13,b14} use Android applications and \cite{b9,b10} use both Android and iOS. The method by which applications are run during data capture periods varies. In \cite{b13}, applications are executed over a fixed time interval. In contrast, \cite{b14} considers two scenarios in which an application is run both with and without background applications. The background applications, however, are not labelled. In \cite{b12}, UI fuzzing techniques were used to automate application use. 

The most relevant dataset to our work is provided in \cite{b11}, which is for classification of VPN and non-VPN application traffic of a single mobile device. Network traffic was manually generated with two user profiles. Seven types of traffic such as web browsing, email, chat, and file transfer were captured along with the application. Our datasets differ from \cite{b11} in a number of key ways. Rather than one device, our datasets consists of three Android and two iOS mobile phones, each with distinct user profiles. For applications, both Gmail and Apple mail are used for email. Spotify and SoundCloud are present in our datasets, whereas \cite{b11} has no music streaming platforms. The key difference, however, between our datasets and \cite{b11} is the automated network traffic and labelling process which we outline in detail in Section \ref{sec: Testbed overivew}. To the best of our knowledge this is the first testbed that meets all the following conditions:

\begin{enumerate}
    \item Not reliant on real-world users.
    \item Evaluates Android and iOS devices.
    \item Automates and labels standard applications without fuzzing, manual labelling or time-interval labelling.
\end{enumerate}

\begin{table*}[t]
\centering
\caption{Example actions of each application on the testbed.}
\begin{tabular}[t]{lll}
\toprule
Application & Action & Description\\
\midrule
Chrome / Safari  & Browser Search & Searches a specific term on a device's native browser. \\
Gmail / Mail & Mail Send & Composes an email and sends it to another device on the testbed\\
Gmail / Mail & Mail Update & Checks for new mail.\\
Messenger / WhatsApp & Message Send & Sends a message to another device on the testbed.\\
Messenger / WhatsApp & Message Update & Checks for new messages.\\
Google / Apple Maps & Map Search & Searches a route between two random points on devices native map application.\\
Spotify / SoundCloud & Search & Searches for a song and plays it.\\
YouTube & YouTube Search & Searches for a video on YouTube watches the video (at least 10 minutes).\\
Instagram / Twitter / Snapchat / Facebook  & Update & Opens application and updates content. \\

\bottomrule
\end{tabular}
\label{table: testbed actions}
\end{table*}%

\section{Testbed overview}
\label{sec: Testbed overivew}

Here we provide an overview of the testbed architecture and the process for automating, capturing, and labelling network traffic.

\subsection{Testbed Topology}
\begin{figure}[t]
\includegraphics[width=\linewidth]{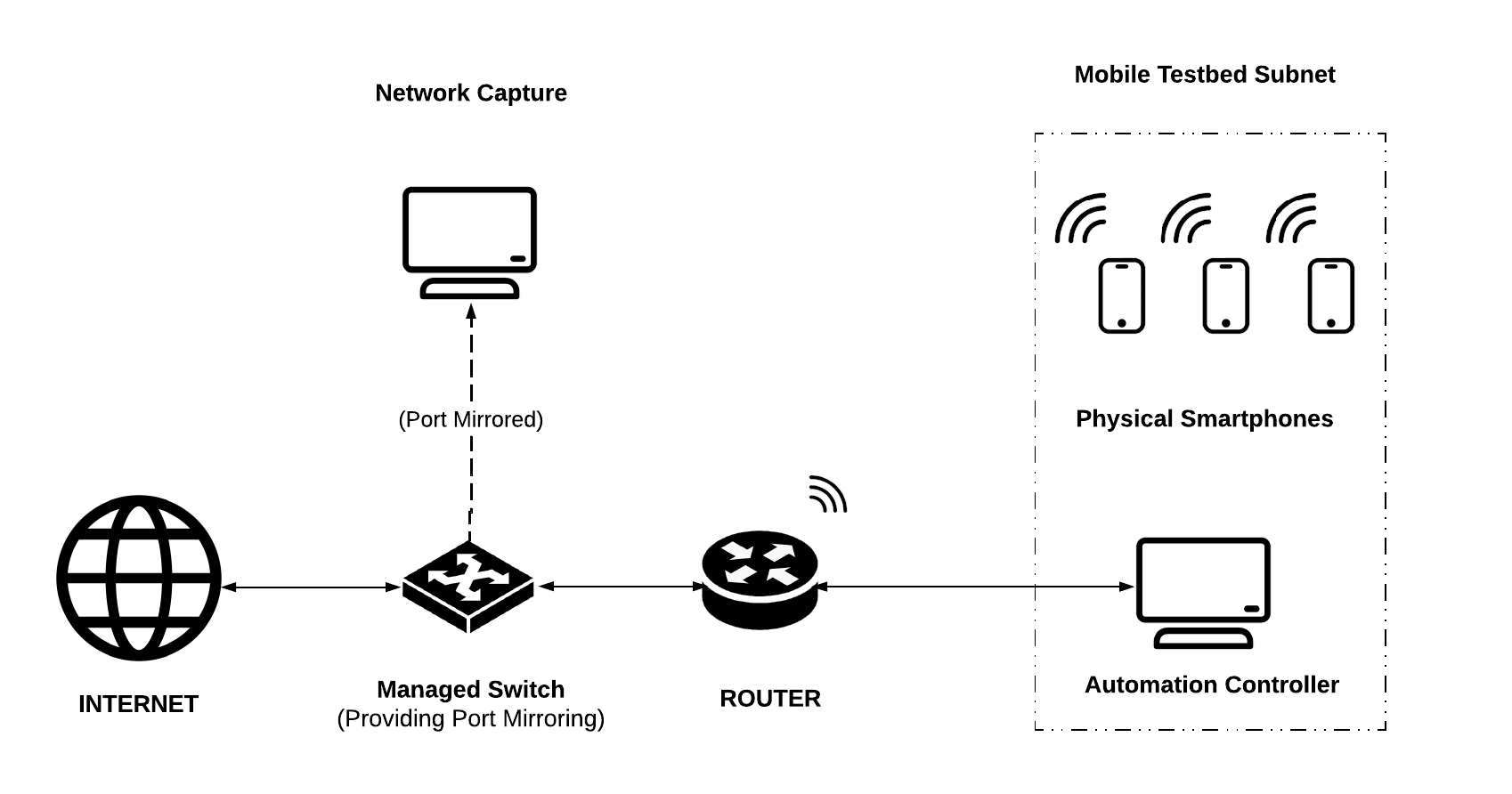}
\caption{Testbed network topology diagram.}
\label{fig:testbed top}
\end{figure}

Figure \ref{fig:testbed top} provides a high-level topology of the testbed network. All devices are connected via USB to the automation controller. To automate iOS devices, an OS X device must be used. As such we have chosen a Mac mini as the automation controller. The Mac mini handles automating applications and processing data. 
We use a dedicated Linux machine for capturing the network traffic. All devices on the testbed are connected to a Linksys wrt1200AC router. The router is connected to a managed switch with port mirroring which captures traffic to and from the router.

\subsection{Application Automation}
Our Python-based automation framework is designed to model realistic user actions of applications. Table \ref{table: testbed actions} provides a list of actions that can be taken by mobile devices. Our automation framework can model a sequence of actions within an application, a key differentiator from UI fuzzing. For example, sending a message to another device on the testbed requires a sequence of actions. This sequence of actions is not guaranteed to occur with UI fuzzing. Datasets created via UI fuzzing are therefore unable to represent realistic application use. 

Appium server\cite{b30}, an open-source client-server framework, is used to automate iOS and Android mobile devices. Appium converts the commands from the Python automation scripts into UIAutomator2 and XCUITest, the automation frameworks for Android and iOS devices respectively. Through UIAutomator2 and XCUITest, Appium server is able to identify and interact with buttons within an application. We utilised this to create sequences of button presses to accurately and efficiently automate user actions. This framework is the first to automate iOS and Android application use without UI fuzzing.


\subsection{Traffic Capture and Labelling}
To start a capture, the automation scripts are run and the packet trace tool tcpdump records all traffic flowing into and out of the testbed subnet. Tcpdump provides a packet capture file (PCAP) containing the packets sent and received during the capture period. Tranalyzer\cite{b32} is then used to process the PCAPs into traffic flows, creating a dataset of network flows from mobile devices. The process of labelling flows on Android and iOS devices differs, so we consider both in turn.

For Android devices, we use the proc file system (procfs) to extract information about the current connections on the device. The key piece of information is the unique identifier (UID) that created the socket connection. The UID remains constant unless the application is reinstalled. We use the Android command dumpsys to link the UID and application. Dumpsys is an Android command providing information about Android system services. 

For iOS devices, we use Apple's remote virtual interface (RVI) with tcpdump to record traffic. Flows are labelled using tcpdump, which obtains the application label from the operating system. Flows are matched to the UID and RVI by their 5-tuple.

During a capture, after an action for an application is executed the application continues running in the background. Unlike time-interval labelling, our process is independent of current application use. We can therefore run and label background application use and not impact the fidelity of our ground-truth. 

\section{Dataset}

The datasets consist of approximately 24,000 labelled flows from two 40-hour captures. The datasets have 36 flow features (20 numerical, 16 categorical) and five mobile devices. Specifically, the testbed has three Android and two iOS devices. We have chosen to extract flow features commonly used in traffic classification, such as protocol information and packet summary statistics. The flow features are separated into local and remote, indicating which device the uni-directional features relates to. The datasets are labelled with application and operating system types, allowing for classification of both the application and operating system. There are 15 mobile applications and two operating systems (iOS and Android) in the datasets. 

It is estimated that 75\% of traffic is encrypted \cite{b24}. Encryption has rendered packet-based traffic classification tools ineffective \cite{23}. Flow-based techniques, however, have been shown to achieve high performance in the presence of encryption. Traffic classification models are therefore predominately flow-based. In line with current research, the datasets provided are of traffic flows rather than PCAPs. 

\begin{figure}[t]
\includegraphics[width=8cm]{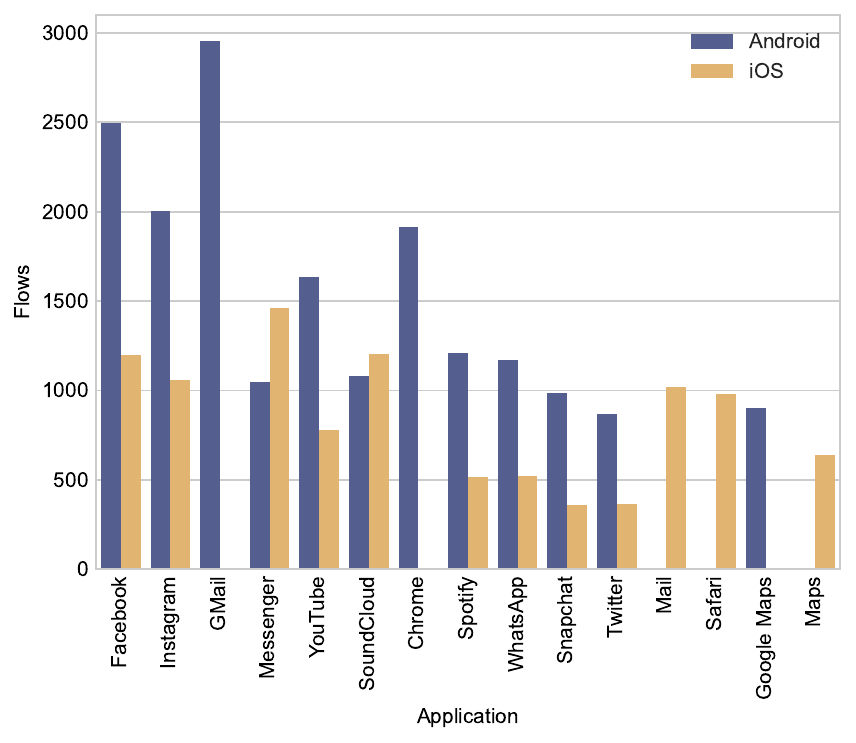}
\caption{Capture 1 distribution of applications by OS.}
\label{fig: OS distribtuion}
\end{figure}

Figure \ref{fig: OS distribtuion} shows the distribution of application by operating system. We see that Gmail, Chrome, and Google Maps flows are only generated by Android devices. Furthermore, Apple Maps, Apple Mail and Safari are only generated by iOS devices. There are several reasons why certain applications are only available on one operating system. Apple Maps and Apple Mail are only available on iOS devices. As such we require a different Android based map and email service. Chrome is only used on Android as Chrome iOS is functionally equivalent to Safari. Both Chrome iOS and Safari use the same browser engine WebKit, however, Chrome iOS uses its own UI.

\begin{figure}[t]
\includegraphics[width=8cm]{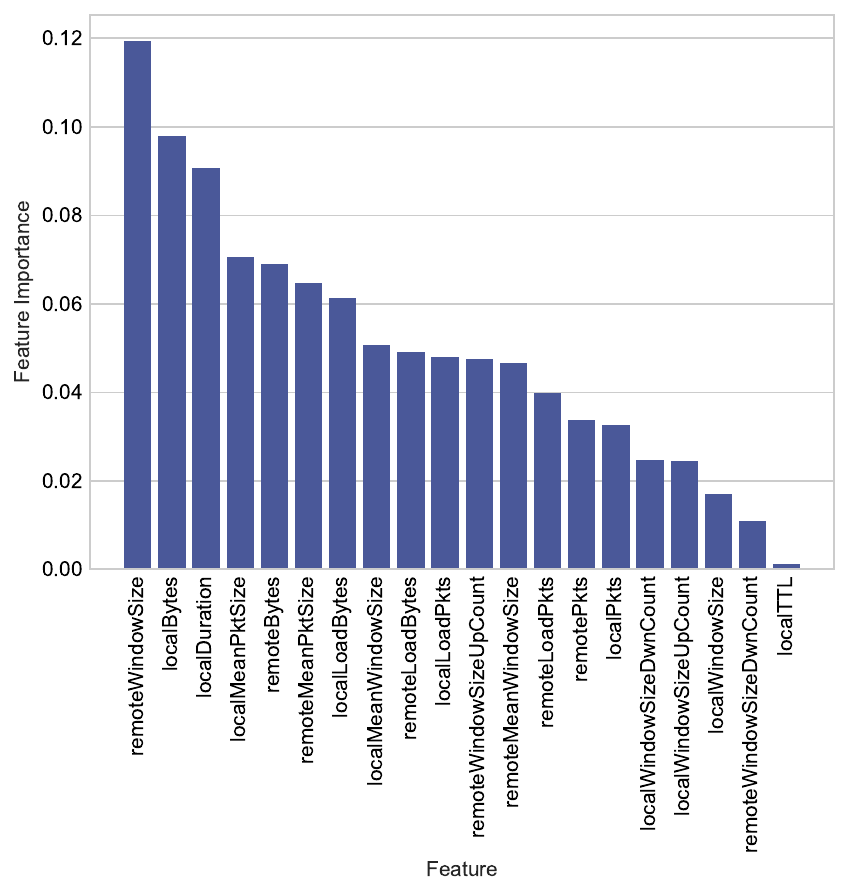}
\caption{Random forest feature importance.}
\label{fig: RF feat importance}
\end{figure}

To evaluate the efficacy of our datasets for application classification, we apply random forest feature selection. We expect that the most influential features for application classification in our datasets are common within the literature. We see in Figure \ref{fig: RF feat importance} that the features of highest importance are statistical summaries of the packets, number of bytes, duration and TCP window size. These features have been experimentally validated in \cite{b11,b8,b17} as beneficial in traffic classification. Furthermore, the features of highest importance are not those set by the operating system such as time-to-live (TTL). This indicates that applications are not exclusively distinguished by their operating system. 

\section{Automation Reliability}

\begin{table}[t]
\centering
\caption{Percentage of failed launches of applications.}
\begin{tabular}[t]{lllll}
\toprule
Application & LF \% (iPhone 6) & LF \% (iPhone 7) & LF \% (Android) \\
\midrule
Mail & $1.25$ & $0.56 $ & 0\\
Facebook  & $0$ & $0.60$ & 0\\
Messenger  & $1.96$ & $0.28 $ & 0\\
Instagram  & $2.41 $ & $0$ & 0\\
SoundCloud  & $3.23 $ & $0$ & 0\\
WhatsApp  & $0.28 $ & $0.33 $ & 0\\
Other & 0 & 0 & 0 \\

\bottomrule
\end{tabular}
\label{table: iPhone fail}
\end{table}%

\begin{table}[t]
\centering
\caption{Percentage of application automation failures.}
\begin{tabular}[t]{lllll}
\toprule
Application & EF \% (iPhone 6) & EF \% (iPhone 7)  & EF \% (Android)\\
\midrule
Twitter & $10$ & $14$ & 0\\
Apple Maps & $1.4$ & $0$ & 0\\
News & $6.7$ & $0$ & 0 \\
YouTube & 0 & 0 & 8\\
SoundCloud & 0 & 0 & 5 \\
Other & 0 & 0 & 0 \\
\bottomrule
\end{tabular}
\label{table: iPhone automation fail}
\end{table}%
In this section, we assess the reliability of the automation process across two 40-hour traffic captures. Two metrics assess reliability, execution failure (EF) and launching failure (LF). Execution refers to the completion of the automation process, while launching refers to the ability of the Appium server to launch an application without crashing. These metrics measure how the Appium server and automation software affect reliability in generating network traffic. We first assess the reliability by application, then by each device on the testbed. 

\subsection{Reliability by Application}

All applications on Android devices launch successfully on both traffic captures. Out of 15 applications on iOS devices, six applications exhibited at least one failed launch. Table \ref{table: iPhone fail} provides the LF percentages for applications on the iPhone 6 and 7. Of the six applications in Table \ref{table: iPhone fail}, four (Facebook, Messenger, Instagram \& WhatsApp) are owned by the parent company Meta. The higher LF and number of iOS applications with a launch failure indicate that Appium has a higher likelihood to fail on iOS then Android.

Two Android applications, YouTube and SoundCloud experienced at least one automation failure with EFs of 8\% and 5\% respectively. In Table \ref{table: iPhone automation fail}, we provide the EF percentages for applications. Twitter is the only application to have an automation failure on both the iPhone 6 and 7. The higher EF rates on the iPhone 6 are due to the device losing connection to the Appium server three times during the first capture. 
The cause for execution failure on Android devices and Twitter on the iPhone 7 have not been determined. These results suggest that iOS versions of applications are more likely to fail to automate and launch. Future work will determine if Appium, different versions of iOS or Meta applications are the cause of a higher EF and LF of applications on iOS devices.

\subsection{Reliability by Device}
Table \ref{table: device rel} shows the reliability metrics for each device. While all Android devices successfully launched their applications, the iPhone 6 and 7 experienced launch failure rates of 1.02\% and 0.288\% respectively. The iOS devices also exhibited a higher proportion of automation failures, with the iPhone 6 suffering an execution failure rate of 0.409\%. As mentioned previously, the iPhone 6's higher EF is due to connection issues with the Appium server. 

Overall, these results suggest that the automation process for each device is highly reliable. Appium on iOS devices, however, has a higher likelihood to fail. It is worth noting that the Huawei device did not experience any failures in launching or execution.
\begin{table}[t]
\centering
\caption{Reliability metrics for each device on the testbed.}
\begin{tabular}[t]{llll}
\toprule
Device & OS & EF \% & LF \% \\
\midrule
Motorola G4 & Android 7	& 0.198  & 0 \\
Samsung S9 & Android 8 & 0.124  &0\\
Huawei & Android 9 & 0  & 0\\
iPhone 6 & iOS 11.3.1  & 0.409 &  1.020\\
iPhone 7 & iOS 13.3.1 & 0.230 & 0.288\\
\bottomrule
\end{tabular}
\label{table: device rel}
\end{table}%

\section{Dataset Benchmarks}
\begin{table}[t]
\centering
\caption{Application classification performance of random forest, SVM and KNN.}
\begin{tabular}[t]{llll}
\toprule
Technique &Macro-Recall & Macro-Precision & Macro-F1\\
\midrule
SVM (RBF) & $0.932$ & $0.931$ & $0.931$ \\
Random Forest  & $0.920$ & $0.956$ & $0.936$\\
KNN (k = 1)  & $0.865$ & $0.854$ & $0.859$\\
\bottomrule
\end{tabular}
\label{table: application classification performance}
\end{table}%

\begin{table}[t]
\centering
\caption{Classification performance of random forest by application.}
\begin{tabular}[t]{lllll}
\toprule
Application & Macro-Recall & Macro-Precision & Macro-F1\\
\midrule
Chrome    & $ 0.926 $  & $ 0.855 $ & $ 0.889 $ \\
Facebook  & $ 0.950 $ & $ 0.920 $ & $ 0.935 $\\
   Gmail  &    0.967   &  0.986   &  0.977   \\
Instagram  &    0.911   &  0.957  &   0.934\\
Apple  Mail &      0.985    & 0.970  &   0.977\\
Apple Maps &     0.973    & 0.985   &  0.979\\
Messenger &     0.991   &  0.988  &   0.990\\
Snapchat   &   0.909    & 0.945   &  0.927\\
 Spotify    &  0.891    & 0.896    & 0.894\\
 Twitter &      0.946   &  0.899  &   0.922\\
WhatsApp  &    0.993    & 0.998  &   0.996\\
 YouTube   &   0.870    & 0.948 &    0.907\\
Safari    &  0.864  &   0.816   &  0.839\\
\bottomrule
\end{tabular}
\label{table: random forest application classification performance}
\end{table}%

To evaluate our datasets in application classification we compared three standard supervised ML algorithms, namely, k-nearest neighbours (KNN), support vector machine (SVM) with radial basis function (RBF) kernel and random forest (RF). The classification performance also provides a benchmark for this dataset, allowing other researchers to compare their techniques. 

Training and testing of classification models are performed on datasets from the first and second capture respectively. Only numerical features were used as it is typical in traffic classification. We leave the use of categorical features to future work. To evaluate the performance of the above classification models we use three common metrics: Macro-Precision, Macro-Recall and Macro-F1. Macro-averaging was chosen as it is robust to class imbalance. The classification performance of each model is given in Table \ref{table: application classification performance}. We see that random forest achieves the highest score followed by SVM and KNN. As these are baseline results, hyper-parameter tuning has not been extensively conducted. 

Table \ref{table: random forest application classification performance} shows the classification performance of random forest on each application. Random forest yielded the lowest performance in classifying web browsers, despite the Chrome application being the largest class. The network characteristics of web browsers are not as distinctive as an application thereby causing lower classification performance.

\section{Limitations of the Testbed}
Here we outline and address the potential limitations in the hardware and automation software of the testbed.

\subsection{Hardware}
One assumption made in creating the testbed is the choice of mobile devices and the versions of operating systems. In a real-world network however, versions of operating systems and applications are constantly changing via updates. Furthermore, at any one time a network will have a diverse mix of devices and operating systems. Maintaining up-to-date software to automate these devices is a challenging task. The mobile devices used in this study were popular at the time the testbed was created. However, manufacturers often release new devices and operating systems each year, making it infeasible to constantly update the hardware on the testbed. Despite this, the testbed automation software is capable of integrating new devices and applications as needed. In the future, emulated versions of mobile devices will be incorporated into the testbed. Emulated devices will enable the testbed to stay up to date with current operating system and application versions.

\subsection{Automation}

The automation process was designed with the assumption that the applications would be used in a uniform way. For example, when automating email, only a set number of lines of text are sent. This does not account for users sending email attachments such as images, videos, and other files. Another example is browsing and watching YouTube, where only the most popular video is selected and watched for a maximum of 10 minutes. This assumption does not necessarily reflect real-world application use. To create effective traffic classification and analysis techniques from this testbed, we require traffic that is realistic. In order to simulate realistic traffic, it is essential that our devices be automated in a way that generates traffic resembling real-world user behaviour. Automating through a set of weighted actions is a successful first approach, but our automation process can be further improved. To improve our automation process, we plan to investigate the use of statistical modelling. Real-world network traffic can be used to create statistical models that provide the basis for how we automate our testbed devices.

\section{Conclusion and Future Work}
In this paper we presented a testbed for automating mobile devices. The traffic generation and labelling process were discussed. The reliability of the automation process was analysed with its limitations being identified. In addition, two datasets generated by the testbed were provided with three standard classifiers in application classification being benchmarked. In future work, we plan to incorporate a VPN into the testbed. Incorporating a VPN into the testbed will enable us to automate and capture VPN traffic, allowing us to study private and encrypted networks. 

\end{document}